\documentclass[11pt]{article}

\usepackage{amsmath, amssymb}
\usepackage{epic,eepic}

\begin{document}

\title{Chromogravity - An Effective Diff(4,R) Gauge for the IR region of
QCD}

\author{Djordje \v Sija\v cki \\
Institute of Physics, P.O. Box 57, 11001 Belgrade, Yugoslavia}

\date{}

\maketitle

\begin{abstract}
Previous work on the IR regime approximation of QCD in which the dominant
contribution comes from a dressed two-gluon effective metric-like field
$G_{\mu\nu} = g_{ab} A^{a}_{\mu} A^{b}_{\nu}$ ($g_{ab}$ a color $SU(3)$
metric) is reviewed. The QCD gauge is approximated by effective
"chromodiffeomorphisms", i.e. by a gauge theory based on a
pseudo-diffeomorphisms group. The second-quantized $G_{\mu\nu}$ field,
together with the Lorentz generators close on the $\overline{SL}(4,R)$
algebra. This algebra represents a spectrum generating algebra for the
set of hadron states of a given flavor - hadronic "manifields"
transforming w.r.t. $\overline{SL}(4,R)$ (infinite-dimensional) unitary
irreducible representations. The equations of motion for the effective
pseudo-gravity are derived from a quadratic action describing Riemannian
pseudo-gravity in the presence of shear ($\overline{SL}(4,R)$ covariant)
hadronic matter currents. These equations yield $p^{-4}$ propagators,
i.e. a linearly rising confining potential $H(r) \sim r$, as well as
linear $J \sim m^{2}$ Regge trajectories.  The $\overline{SL}(4,R)$
symmetry based dynamical theory for the QCD IR region is successfully
applied to hadron resonances. The pseudo-gravity potential reaches over
to Nuclear Physics, where its $J^{P} = 2^{+},\ 0^{+}$ quanta provide for
the ground state excitations of the Arima-Iachello Interacting Boson
Model.
\end{abstract}

\newpage

\section{Introduction}

One of the main challenges in Particle Physics is the understanding
and/or classification of quite a large number of presently known hadronic
resonances. Here we are faced with an intriguing situation: In the
"horizontal" direction one has flavor symmetries and rather powerful
quantitative techniques with practically none understanding of the
corresponding underlying fundamental interaction. As for the "vertical"
direction (fixed flavor content), the basic interaction is given by the
presently widely accepted Quantum Chromodynamics (QCD) theory, however
the non-perturbative features of QCD have made it difficult to apply the
theory exactly. Quite a number of approaches to deal with this region
have been proposed so far with different degree of success. We believe
that the merits of the approach described in this paper are both the fact
that our starting point is QCD itself and that the predictions fit very
well with experiment.

If the hadron lowest ground states are colorless (our assumption) and in
the approximation of an external QCD potential, the hadron spectrum above
these levels will be generated by color-singlet quanta, whether made of
dressed two-gluon configurations, three-gluons, \dots . Every possible
configuration will appear. No matter what the mechanism responsible for a
given flavor state, the next vibrational, rotational or pulsed excitation
corresponds to the "addition" of one such collective color-singlet
multigluon quantum superposition. In the fully relativistic QCD theory,
these contributions have to come from summations of appropriate Feynman
diagrams, in which dressed $n$-gluon configurations are exchanged. We
rearrange the sum by lumping together contributions from $n$-gluon
irreducible parts, $n=2, 3, ..., \infty$ and with the same Lorentz
quantum numbers. The simplest such system will have the quantum numbers
of di-gluon, i.e. $n=2$. The color singlet external field can thus be
constructed from the QCD gluon field as a sum ($g_{ab}$ is the color-$SU(3)$
metric, $d_{abc}$ are the totally symmetric $8\otimes 8\otimes
8\rightarrow 1$ coefficients)
\begin{equation}
g_{ab}A^{a}_{\mu}A^{b}_{\nu} \oplus
d_{abc}A^{a}_{\mu}A^{b}_{\nu}A^{c}_{\sigma} \oplus \cdots .
\end{equation}
In the above, $A^{a}_{\mu}$ is the dressed gluon field.

\section{Chromometric $G_{\mu\nu}$}

We suggest that the main feature of hadron excitations is due to a
component of QCD representing the exchange of a two-gluon effective
gravity-like "chromo-metric" field ($A^{a}_{\mu}(x)$ the properly
normalized gluon) \cite{chromo}:
\begin{equation}
G_{\mu\nu}(x) = g_{ab} A^{a}_{\mu}A^{b}_{\nu} .
\end{equation}

It will be useful for the applications to separate the "flat connection"
$N^{a}_{\mu}$, i.e. the zero-mode of the field. Writing for the curvature
or field strength
\begin{equation}
F^{a}_{\mu\nu} = \partial_{\mu}A^{a}_{\nu} -
\partial_{\nu}A^{a}_{\mu} - if^{a}_{\ \ bc}A^{b}_{\mu}A^{c}_{\nu},
\end{equation}
we define
\begin{equation}
A^{a}_{\mu} = N^{a}_{\mu} + B^{a}_{\mu}, \quad
\partial_{\mu}N^{a}_{\nu} - \partial_{\nu}N^{a}_{\mu} =
if^{a}_{bc}N^{b}_{\mu}N^{c}_{\nu},
\end{equation}
\noindent where
$N^{a}_{\mu}$ is the constant component, yielding a vanishing field
strength.

Such a vacuum solution might be of the instanton type, for instance.
Consider, e.g. the first nontrivial class, with Pontryagin index $n=0$.
Expand around this classical configuration, working, as always for
instantons, in a Euclidean metric (i.e. a tunneling solution in
Minkowski spacetime). At large distances the instanton field is required
to approach a constant value
\begin{equation}
g_{ab}\ N^a_\mu\ \partial_\nu\epsilon^b=\partial_\nu(g_{ab}\ N^a_\mu
\ \epsilon^b)
\end{equation}
with the $B^a_\mu (x)$ field representing a fluctuation around the
constant value, vanishing at large distances. One can construct the
constant vacuum solution by mapping $SU(3)\to S^4$, namely directly
onto the complete Euclidean manifold, compactified by the addition of
a point at infinity.

$G_{\mu\nu}$ acts as a "pseudo-metric" field, (passively)
gauging effective "pseudo-diffeomorphisms", just as is done by the
physical Einstein metric field for the "true" diffeomorphisms of the
covariance group.

The variation of the chromo-metric under color-$SU(3)$, due to
\begin{equation}
{\delta}_{\epsilon} A^{a}_{\mu} = {\partial}_{\mu}{\epsilon}^{a} +
A^{b}_{\mu} ({\lambda}_{b})^{a}_{c} {\epsilon}^{c},
\end{equation}
(we use the adjoint representation
$\{\lambda_b\}^{a}_{c} = -if_{b\ c}^{a} = if^{a}_{bc}$) reads
\begin{eqnarray*}
{\delta}_{\epsilon}G_{\mu\nu} &=& {\delta}_{\epsilon} \{ g_{ab}
(N^{a}_{\mu} + B^{a}_{\mu}) (N^{b}_{\nu} + B^{b}_{\nu}) \} \\
&=& g_{ab}
({\partial}_{\mu}{\epsilon}^{a} N^{b}_{\nu} + N^{a}_{\mu}{\partial}_{\nu}
{\epsilon}^{b} + {\partial}_{\mu}{\epsilon}^{a} B^{b}_{\nu} +
B^{a}_{\mu}{\partial}_{\nu}{\epsilon}^{b}) \\
&&+ ig_{ab} \left\{ f^a_{cd} A^c_\mu \epsilon^d A^b_\nu
+ f^b_{cd} A^a_\mu A^c_\nu \epsilon^d \right\} .
\end{eqnarray*}

The last bracket vanishes, since it represents the homogeneous $SU(3)$
transformation of the $SU(3)$ scalar expression, i.e.
\begin{equation*}
if_{bcd}\ (A^b_\mu\ A^c_\nu + A^c_\mu\ A^b_\nu)\ \epsilon^d
\end{equation*}
(or, more technically, due to the total antisymmetry of $f_{abc}$ in a
compact group).

We note that at the IR region distances, any Gauss theorem
field-fluxes will only involve the $N^a_\mu$ constant component,
whereas the $B^a_\mu(x)$ ``fluctuation"
will not contribute.  As a result, when integrating by parts the terms
in $B^a_\mu,\ B^b_\nu$ we get
\begin{equation*}
g_{ab}\ (\epsilon^a\ \partial_\mu B^b_\nu+\partial_\nu B^a_\mu\ \epsilon^b)
\end{equation*}
an expression whose Fourier transform vanishes for $k \to 0$, i.e. in the
infrared sector. A generalized definition of
this ``IR limit" will be addressed below.

The terms involving the constant $N^{a}_{\mu}$, $N^{b}_{\nu}$ can be
rewritten in terms of effective pseudo-diffeomorphisms, defined by
\begin{equation}
{\xi}_{\mu} \equiv g_{ab} {\epsilon}^{a} N^{b}_{\mu}, \quad
{\delta}_{\epsilon}G_{\mu\nu} = {\partial}_{\mu}{\xi}_{\nu} +
{\partial}_{\nu}{\xi}_{\mu}.
\end{equation}

{\it Thus, the local $SU(3)$ color gauge variations contain a subsystem
ensuring that the $G_{\mu\nu}$ di-gluon indeed act as a "pseudo-metric"
field, precisely emulating gravity}.

The definition of our "IR limit" based on the vanishing of the
$4$-momenta of the `fluctuating fields' $B^{a}_{\mu}$ -- after an
integration by parts in which only the constant fields $N^{a}_{\mu}$
contribute to the surface terms -- will be extended so as to include
similar terms with vanishing momenta in all many-gluon zero-color
exchanges. This can be taken as an operational definition, sufficient for
our general purpose. To gain some additional insight, however, we remind
the reader that such an IR approximation of QCD can also be thought of as
the first step, the "zeroth approximation", of a {\it strong coupling}
regime -- in terms of a "small parameter" representing the number of
"hard", or nonsoft, virtual quanta held in the evaluation of any physical
quantity.  We can write a generic IR state, carrying $4$-momentum $k$, as
follows:
\begin{equation}
\vert \phi_{IR},\ k\rangle = \sum_{m=1}^{\infty}
f_{m}(k_{1},k_{2},\dots ,k_{m})
\delta_{k,k_{1}+k_{2}+\cdots +k_{m}}
\vert k_{1}k_{2}\dots k_{m} \rangle
\end{equation}
where $\vert k_{1}k_{2}\dots k_{m} \rangle$ represents a state of $m$
soft gluons ($k_{i} \approx 0,\ i=1,2,\dots m$).  Integrating by parts
(with surface terms again appearing only for the constant parts), the
matrix elements of the terms in $B^{a}_{\mu}$, $B^{b}_{\nu}$ become in
this IR approximation
\begin{equation*}
\langle \phi_{IR}^{\prime}, k^{\prime}\vert
g_{ab} (\epsilon^{a} \partial_{\mu} B^{b}_{\nu} +
\partial_{\nu} B^{a}_{\mu} \epsilon^{b})
\vert \phi_{IR},\ k\rangle ,
\end{equation*}
an expression that is proportional to the soft $1$-gluon momentum, and
that vanishes for $k\rightarrow 0$, i.e., in the infrared sector. As a
result, when changing over to the $\xi_{\mu}$ variable of and
reidentifying $\delta_{\xi}$ as a variation under a formal $R^{4}$
diffeomorphism, we get ${\delta}_{\epsilon}G_{\mu\nu} =
{\partial}_{\mu}{\xi}_{\nu} + {\partial}_{\nu}{\xi}_{\mu}$. For the
sake of completeness, we note that in general one has to consider
expressions of the following form
\begin{equation*}
\langle \phi_{IR}^{\prime}, k^{\prime}\vert
O(B^{a}_{\mu}, \partial_{\nu}B^{a}_{\mu}) \delta_{\epsilon} G_{\mu\nu}
\vert \phi_{IR},\ k\rangle .
\end{equation*}
We evaluate such expressions, in this IR approximation, by inserting a
complete set of states, and retaining only the {\it soft} virtual
quanta. It is explained in Ref. \cite{Fried} that by making use of the
Fradkin representation \cite{Fradkin} for relevant Green's functions
one has a continuous family of "soft", or IR approximations, which
maintain gauge invariance. Thus, one finds a consistent
gauge-invariant (strong coupling) IR approximation with dressed gluon
propagators which incorporate the iteration of all relevant quark
bubbles, each carrying all possible internal, soft-gluon lines.

The consistency of this IR approximation requires one to consider only
those QCD variations that connect IR gluon configurations {\it
mutually}. Let us consider the expression for the $B=A-N$ variation, i.e.
$\delta_{\epsilon} A^{a}_{\mu} = \partial_{\mu}\epsilon^{a} + i
f^{a}_{\ bc} A^{b}_{\mu} \epsilon^{c}$
The left hand side of this expression is a difference between two soft
gluons, implying that the IR matrix elements of its partial derivative
are soft. Thus, we find the following "IR constraint" on the QCD gauge
parameters:
\begin{equation}
\langle \phi_{IR}^{\prime}, k^{\prime}\vert
\partial_{\rho} \partial_{\mu} \epsilon^{a} +
i f^{a}_{\ bc}  B^{b}_{\mu} \partial_{\rho} \epsilon^{c}
\vert \phi_{IR},\ k\rangle
\approx 0 .
\end{equation}

\section{$Diff(4,R)$ Structure -- $n$-gluon fields}

Let us now consider the multi-gluon colorless configurations \cite{diff}.
The color-singlet $n$-gluon field operator has the following form
\begin{equation}
G^{(n)}_{\mu_{1}\mu_{2}\cdots\mu_{n}} = d^{(n)}_{a_{1}a_{2}\cdots a_{n}}
A^{a_{1}}_{\mu_{1}}A^{a_{2}}_{\mu_{2}} \cdots A^{a_{n}}_{\mu_{n}}
\end{equation}
where
\begin{eqnarray*}
&&d^{(2)}_{a_{1}a_{2}} = g_{a_{1}a_{2}}, \\ &&d^{(3)}_{a_{1}a_{2}a_{3}} =
d_{a_{1}a_{2}a_{3}}, \\ &&d^{(n)}_{a_{1}a_{2}\cdots a_{n}} =
d_{a_{1}a_{2}b_{1}} g^{b_{1}c_{1}} d_{c_{1}b_{2}a_{3}} \cdots \\ &&\hskip
50pt \times g^{b_{n-4}c_{n-4}} d_{c_{n-4}b_{n-3}a_{n-2}}
g^{b_{n-3}c_{n-3}} d_{c_{n-3}a_{n-1}a_{n}},\quad n>3,
\end{eqnarray*}
$A^{a}_{\mu}$ is the dressed gluon field, $g_{a_{1}a_{2}}$ is the $SU(3)$
Cartan metric, and $d_{a_{1}a_{2}a_{3}}$ is the $SU(3)$ totally symmetric
$8\times 8\times 8 \rightarrow 1$ tensor.

But taking Fourier transforms -- i.e. the matrix elements for these gluon
fluctuations -- we find that {\it these terms are precisely those that
vanish in our definition of an IR region}. The terms involving the
constant connections $N^{a_{i}}_{\mu_{i}}$, $i=1,2,\dots n$ can be
rewritten in terms of effective pseudo-diffeomorphisms,

The QCD variation, in the IR region can be rewritten in terms of
effective pseudo-diffeomorphisms,
\begin{equation}
\delta_{\epsilon} G^{(n)}_{\mu_{1}\mu_{2}\cdots\mu_{n}} =
\partial_{\{ \mu_{1}} \xi^{(n-1)}_{\mu_{2}\mu_{3}\cdots\mu_{n}\} }
\equiv \delta_{\xi} G^{(n)}_{\mu_{1}\mu_{2}\cdots\mu_{n}},
\end{equation}
where $\{ \mu_{1}\mu_{2}\cdots\mu_{n}\}$ denotes symmetrization of
indices,
\begin{equation}
\xi^{(n-1)}_{\mu_{1}\mu_{2}\cdots\mu_{n-1}} \equiv
d^{(n)}_{a_{1}a_{2}\cdots a_{n}} N^{a_{1}}_{\mu_{1}}N^{a_{2}}_{\mu_{2}}
\cdots N^{a_{n-1}}_{\mu_{n-1}}\epsilon^{a_{n}}
\end{equation}
while $N^{a_{i}}_{\mu_{i}}$, $i=1,2,\dots n$, being the constant
connections.

A subsequent application of two $SU(3)$-induced variations, i.e. the
commutator of two such chromo-diffeomorphic variations
\begin{equation}
[\delta_{\epsilon_{1}},\ \delta_{\epsilon_{2}}]
G^{(n)}_{\mu_{1}\mu_{2}\cdots\mu_{n}} = \delta_{\epsilon_{3}}
G^{(n)}_{\mu_{1}\mu_{2}\cdots\mu_{n}},
\end{equation}
i.e.
\begin{equation}
[\delta_{\xi_{1}},\ \delta_{\xi_{2}}]
G^{(n)}_{\mu_{1}\mu_{2}\cdots\mu_{n}} = \delta_{\xi_{3}}
G^{(n)}_{\mu_{1}\mu_{2}\cdots\mu_{n}},
\end{equation}
where
\begin{equation}
\xi_{3\mu} = (\partial_\nu\xi_{1\mu})\ \xi_2^\nu +
(\partial_\mu\xi_{1\nu})\ \xi_2^\nu - (\partial_\nu\xi_{2\mu})\
\xi_1^\nu - (\partial_\mu\xi_{2\nu})\ \xi_1^\nu
\end{equation}
indeed closes on the covariance group's commutation relations.
Thus, one has an infinitesimal nonlinear realization of the $Diff(4,R)$
group in the space of fields $\Big\{
G^{(n)}_{\mu_{1}\mu_{2}\cdots\mu_{n}}\ \Big\vert \ n=2,3,\dots \Big\}$.

\section{$Diff(4,R)$ Structure --  $L^{(m)}$ Operators Algebra}

Let us consider an $\infty$-dimensional vector space over the field
operators \break\hfill
$\Big\{ G^{(n)} \ \Big\vert\ n=2,3,\dots \Big\}$,
i.e.,
\begin{equation}
V(G^{(2)},G^{(3)},\dots )\ =\ V( G^{(2)}_{\mu_{1}\mu_{2}},\
G^{(3)}_{\mu_{1}\mu_{2}\mu_{3}},\dots ).
\end{equation}
We can now define an infinite set of field-dependent operators \break\hfill
$\Big\{ L^{(m)}\ \Big\vert\ m=0,1,2,\dots \Big\}$ as follows

\begin{eqnarray*}
&&L^{(0)\rho}_{\nu_{1}}\ =\ d^{(2)}_{a_{1}a_{2}}A^{a_{1}}_{\nu_{1}}
\frac{\delta}{\delta (g_{a_{2}b}A^{b}_{\rho})}\ \equiv\ g_{a_{1}a_{2}}
A^{a_{1}}_{\nu_{1}} \frac{\delta}{\delta (g_{a_{2}b}A^{b}_{\rho})}, \\
&&L^{(1)\rho}_{\nu_{1}\nu_{2}}\ =\ d^{(3)}_{a_{1}a_{2}a_{3}}
A^{a_{1}}_{\nu_{1}}A^{a_{2}}_{\nu_{2}} \frac{\delta}{\delta
(g_{a_{3}b}A^{b}_{\rho})}\ \equiv\ d_{a_{1}a_{2}a_{3}}
A^{a_{1}}_{\nu_{1}}A^{a_{2}}_{\nu_{2}} \frac{\delta}{\delta
(g_{a_{3}b}A^{b}_{\rho})}, \\ &&\cdots\cdots \\
&&L^{(m)\rho}_{\nu_{1}\nu_{2}\cdots\nu_{m+1}}\ =\
d^{(m+2)}_{a_{1}a_{2}\cdots a_{m+2}}
A^{a_{1}}_{\nu_{1}}A^{a_{2}}_{\nu_{2}} \cdots A^{a_{m+1}}_{\nu_{m+1}}
\frac{\delta}{\delta (g_{a_{m+2}b}A^{b}_{\rho})}. \\ &&\cdots\cdots .
\end{eqnarray*}

In the general case, $L^{(m)\rho}_{\nu_{1}\nu_{2}\cdots\nu_{m+1}}$,
$m=0,1,2,\dots $ action on the field operators \hfill\break $\Big\{
G^{(n)}\ \Big\vert\ n=2,3,\dots \Big\}$ reads

\begin{eqnarray*}
&&L^{(m)\rho}_{\nu_{1}\nu_{2}\cdots\nu_{m+1}} G^{(2)}_{\mu_{1}\mu_{2}}\
=\ \delta^{\rho}_{\mu_{1}}
G^{(2+m)}_{\nu_{1}\nu_{2}\cdots\nu_{m+1}\mu_{2}} +
\delta^{\rho}_{\mu_{2}} G^{(2+m)}_{\mu_{1}\nu_{1}\nu_{2}\cdots\nu_{m+1}},
\\ &&L^{(m)\rho}_{\nu_{1}\nu_{2}\cdots\nu_{m+1}}
 G^{(3)}_{\mu_{1}\mu_{2}\mu_{3}}\ =\
\delta^{\rho}_{\mu_{1}}
G^{(3+m)}_{\nu_{1}\nu_{2}\cdots\nu_{m+1}\mu_{2}\mu_{3}} +
\delta^{\rho}_{\mu_{2}}
G^{(3+m)}_{\mu_{1}\nu_{1}\nu_{2}\cdots\nu_{m+1}\mu_{3}} \\ &&\hskip120pt
+ \delta^{\rho}_{\mu_{3}}
G^{(3+m)}_{\mu_{1}\mu_{2}\nu_{1}\nu_{2}\cdots\nu_{m+1}}, \\
&&\cdots\cdots \\ &&L^{(m)\rho}_{\nu_{1}\nu_{2}\cdots\nu_{m+1}}
 G^{(n)}_{\mu_{1}\mu_{2}\cdots\mu_{n}}\ =\
\delta^{\rho}_{\mu_{1}}
G^{(n+m)}_{\nu_{1}\nu_{2}\cdots\nu_{m+1}\mu_{2}\cdots\mu_{n}} +
\delta^{\rho}_{\mu_{2}}
G^{(n+m)}_{\mu_{1}\nu_{1}\nu_{2}\cdots\nu_{m+1}\mu_{3}\cdots\mu_{n}} \\
&&\hskip120pt + \cdots + \delta^{\rho}_{\mu_{n}}
G^{(n+m)}_{\mu_{1}\mu_{2}\cdots\mu_{n- 1}\nu_{1}\nu_{2}\cdots\nu_{m+1}},
\\ &&\cdots\cdots .
\end{eqnarray*}

Let us now consider the algebraic structure defined by the \break\hfill
$\Big\{ L^{(m)}\ \Big\vert\ m=0,1,2,\dots \Big\}$ operators Lie
brackets. For the $L^{(0)}$ operators themselves we find
\begin{equation}
[L^{(0)}, L^{(0)}] \subset L^{(0)},
\end{equation}
i.e.
\begin{equation}
[L^{(0)\rho_{1}}_{\nu_{1}}, L^{(0)\rho_{2}}_{\sigma_{1}}] =
\delta^{\rho_{1}}_{\sigma_{1}}L^{(0)\rho_{2}}_{\nu_{1}} -
\delta^{\rho_{2}}_{\nu_{1}}L^{(0)\rho_{1}}_{\sigma_{1}},
\end{equation}

In the most general case, for the brackets of $L^{(l)}$ and $L^{(m)}$ we
find
\begin{equation}
[L^{(l)}, L^{(m)}] \subset L^{(l+m)},
\end{equation}
and more specifically,
\begin{eqnarray*}
[L^{(l)\rho_{1}}_{\nu_{1}\nu_{2}\cdots\nu_{l+1}},
L^{(m)\rho_{2}}_{\sigma_{1}\sigma_{2}\cdots\sigma_{m+1}}] =&&
\sum_{i=1}^{m+1} \delta^{\rho_{1}}_{\sigma_{i}}
L^{(l+m)\rho_{2}}_{\sigma_{1}\sigma_{2}\cdots\sigma_{i-1}
\nu_{1}\nu_{2}\cdots\nu_{l+1}\sigma_{i+1}\cdots\sigma_{m+1}}\\ 
&-& \sum_{j=1}^{l+1} \delta^{\rho_{2}}_{\nu_{j}}
L^{(l+m)\rho_{1}}_{\nu_{1}\nu_{2}\cdots\nu_{j-1}\sigma_{1}\sigma_{2}
\cdots\sigma_{m+1}\nu_{j+1}\cdots\nu_{l+1}}.
\end{eqnarray*}

We have constructed an $\infty$-component vector space, $V$ $=$ \break
$V( G^{(2)}_{\mu_{1}\mu_{2}},\ G^{(3)}_{\mu_{1}\mu_{2}\mu_{3}}, \dots )$,
over the $n$-gluon field operators, as well as the corresponding algebra
of homogeneous diffeomorphisms,
\begin{equation}
diff_{0}(4,R)\ =\ \Big\{ L^{(m)\rho}_{\nu_{1}\nu_{2}\cdots\nu_{m+1}}
\Big\vert m=0,1,2, \dots \Big\};
\end{equation}
the vector space $V$ is invariant
under the action of the $diff_{0}(4,R)$ algebra.

Let us point out that there exists a subalgebra of the entire algebra
when $m$-values are even, i.e. one has the following structure
\begin{eqnarray*}
&&[L^{(even)}, L^{(even)}] \subset L^{(even)}, \\ 
&&[L^{(even)}, L^{(odd)}] \subset L^{(odd)}, \\ 
&&[L^{(odd)}, L^{(odd)}] \subset L^{(even)}.
\end{eqnarray*}

Let us define the dilation-like operator (chromo-dilation) $D$ as a trace
of $L^{(0)\rho}_{\nu}$, i.e.,
\begin{equation}
D = L^{(0)\rho}_{\rho}.
\end{equation}
This operator commutes with the $L^{(0)\rho}_{\nu}$ operators,
\begin{equation}
[D, L^{(0)\rho}_{\nu}] = 0,
\end{equation}
and belongs to the center of the $gl(4,R)$ chromo-gravity subalgebra
generated by the $L^{(0)\rho}_{\nu}$ operators. On account of the
chromo-dilation operator one can make the following decomposition
\begin{equation}
gl(4,R) = r \oplus sl(4,R),
\end{equation}
where $D$ corresponds to the subalgebra $r$, while the basis of the
$sl(4,R)$ subalgebra is given by
\begin{equation}
T^{(0)\rho}_{\nu} = L^{(0)\rho}_{\nu} -
\hbox{$\frac{1}{4}$}\delta^{\rho}_{\nu}D.
\end{equation}

The commutation relation of $D$ with a generic $diff_{0}(4,R)$ operator
$L^{(m)\rho}_{\nu_{1}\nu_{2}\cdots\nu_{m+1}}$ reads
\begin{equation}
[D, L^{(m)\rho}_{\nu_{1}\nu_{2}\cdots\nu_{m+1}}] = m
L^{(m)\rho}_{\nu_{1}\nu_{2}\cdots\nu_{m+1}}
\end{equation}
and thus, the chromo-dilation operator $D$ provides us with a $Z_{+}$
grading. This grading justifies and/or explains the $m$-label used for
the $L^{(m)\rho}_{\nu_{1}\nu_{2}\cdots\nu_{m+1}}$ operators.

The chromo-dilation operator $D$ counts the number of single gluon fields
in a multi-gluon configuration, as seen from the following commutation
relation
\begin{equation}
[D, G^{(n)}_{\mu_{1}\mu_{2}\cdots\mu_{n}}] = n
G^{(n)}_{\mu_{1}\mu_{2}\cdots\mu_{n}}.
\end{equation}

Clearly, the $J=1$ Yang-Mills gauge of QCD contains (in the IR limit)
local diffeomorphisms, gauged \`a la Einstein. As a matter of fact, this
is not surprising, as the truncated massless sector of the open string
reduces to a $J=1$ Yang-Mills field theory and that the same truncation
for the closed string reduces to a $J=2$ gravitational field theory; but
the closed string is nothing but the contraction of two open strings!

\section{Chromogravity Matter Fields}

The effective QCD interaction fields $G_{\mu\nu}$  couple to the hadron
systems themselves, and thus, in order to complete the Chromogravity
approximation of the QCD IR region, we have to address the question of
the effective hadron fields as well. It is well known that the
constructions of hadrons, i.e. the  composite objects made of quarks and
gluons, is due to the strong coupling regime one of the most challenging
issues in QCD. Thus, in order to define the effective hadron fields of
Chromogravity we rely as much as possible on the symmetry considerations.

The $G_{\mu\nu}$ fields, that transform w.r.t. the second-rank symmetric
representation of the $Diff(4,R)$ group, are naturally coupled to the
bosonic and fermionic hadron fields that transform themselves w.r.t.
representations of the $Diff(4,R)$ group as well \cite{wspin}.

The construction of the fermionic fields requires the study of the
quantum-mechanical $Diff(4,R)$ group, i.e. $Diff(4,R)_{QM}$. Note that
the topological properties of the $Diff(4,R)$ group that determine
nontrivial minimal group-extensions of the $Diff(4,R)$ group by the
$U(1)$ group of the quantum-mechanical Hilbert space phase factors
\begin{eqnarray}
1 \rightarrow U(1) \rightarrow Diff(4,R)_{QM}  \rightarrow
Diff(4,R) \rightarrow 1
\end{eqnarray}
are given by the corresponding properties of the group chain:
\begin{eqnarray}
Diff(4,R) \supset GL(4,R) \supset SL(4,R) \supset SO(3,1) \supset SO(3) .
\end{eqnarray}

It is well known that, in contradistinction to $SO(3,1)$ and $SO(3)$
cases, the $SL(4,R)$ group cannot be embedded into any group of finite
complex matrices, and that the universal covering of the $SL(4,R)$ group,
i.e. $\overline{SL}(4,R)$, is a group of infinite matrices -- likewise
for the $\overline{Diff}(4,R)$. The universal covering is actually given
by the double covering, and the corresponding relations among relevant
symmetry groups are as presented in the following diagram:

\begin{center}
\begin{tabular}{ccccccccc}
$1$ & $\rightarrow$ & $Z_2$ & $\rightarrow$ & $\overline{Diff}(4,R)$ &
$\rightarrow$ & $Diff(4,R)$ & $\rightarrow$ & $1$ \\
 & & & & $\cup$ & & $\cup$ & & \\
$1$ & $\rightarrow$ & $Z_2$ & $\rightarrow$ & $\overline{SL}(4,R)$ &
$\rightarrow$ & $SL(4,R)$ & $\rightarrow$ & $1$ \\
 & & & & $\cup$ & & $\cup$ & & \\
$1$ & $\rightarrow$ & $Z_2$ & $\rightarrow$ & $\overline{SO}(3,1)$ &
$\rightarrow$ & $SO(3,1)$ & $\rightarrow$ & $1$ \\
 & & & & $\cup$ & & $\cup$ & & \\
$1$ & $\rightarrow$ & $Z_2$ & $\rightarrow$ & $\overline{SO}(3)$ &
$\rightarrow$ & $SO(3)$ & $\rightarrow$ & $1$ \\
\end{tabular}
\end{center}

An immediate consequence is that there are no finite-dimensional
spinorial representations of the $\overline{SL}(4,R)$, i.e.
$\overline{Diff}(4,R)$ group -- {\em all unitary and non-unitary
spinorial representations of these groups are necessarily
infinite-dimensional}. In practice, the $\overline{SL}(4,R)$
representations are constructed by making use of the "standard" linear
representations techniques, while the $\overline{Diff}(4,R)$
representations are induced from these $\overline{SL}(4,R)$
representations. This fact fits very well with our Chromogravity picture
of hadrons, where the entire set of presumably infinitely many
excitations of given flavor are to be described by a single
infinite-component effective field -- "manifield''.

In order to set up all basic quantum mechanical objects, that are
necessary for particle physics applications, we have to consider:

(i) The fermionic and bosonic (infinite-component) representations of the
$\overline{SL}(4,R)$ group that characterize respectively the baryonic
and mesonic quantum manifields,

(ii) The fermionic and bosonic (infinite-component) representations of
the inhomogenious $SA(4,R) = T_4 \wedge \overline{SL}(4,R)$ group (affine
generalization of the Poincar\'e group) that characterize the quantum
states of the manifields quanta,

(iii) The wave equation type relations that insure consistency between
manifields and the corresponding quantum states, and

(iv) The physical requirements that are primarily related to the
unitarity properties of various observable facts.

The affine group $\overline{SA}(4,R) = T_{4} \wedge \overline{SL}(4,R)$,
is a semidirect product of translations generated by $P_{\mu},\ \mu =
0,1,2,3$ and $\overline{SL}(4,R)$ generated by $Q_{\mu\nu}$ ($\mu ,\nu
=0,1,2,3$). The antisymmetric operators $M_{\mu\nu} = \frac{1}{2}
(Q_{\mu\nu} - Q_{\nu\mu})$ generate the Lorentz subgroup
$\overline{SO}(3,1)$, while the symmetric traceless operators (shears)
$T_{\mu\nu} = \frac{1}{2} (Q_{\mu\nu} + Q_{\mu\nu}) - \frac{1}{4}
\eta_{\mu\nu} Q_{\sigma}^{\ \sigma}$ generate the proper
$4$-volume-preserving deformations.

As in the Poincar\'e case, the $\overline{SA}(4,R)$ unitary
irreducible representations are induced from the representations of
the 
corresponding little group $T^{\prime}_{3}
\wedge \overline{SL}(3,R),\ m\neq 0$. In the physically most interesting
case $T^{\prime}_{3}$ is represented trivially. The corresponding
particle states have to be described by the unitary representations of
the remaining part of the little group, i.e. $\overline{SL}(3,R)$. All
these representations, both spinorial and tensorial, are
infinite-dimensional owing to the $\overline{SL}(3,R)$ noncompactness.
Therefore, the corresponding $\overline{SL}(4,R)$ matter fields $\Psi
(x),\ \Phi (x)$ are necessarily infinite-dimensional and when reduced
with respect to the $\overline{SL}(3,R)$ subgroup should transform with
respect to its unitary irreducible quantum-states representations .

If the whole $\overline{SL}(4,R)$ group would be represented unitarily,
the Lorentz boost generators intrinsic part would be hermitian and as a
result, when boosting a particle, one would obtain a particle with a
different spin, i.e. another particle - contrary to experience. There
exists however a remarkable inner {\it deunitarizing} automorphism ${\cal
A}$ \cite{sl4} of the $SL(4,R)$ group, which leaves its $R_{+} \otimes
\overline{SL}(3,R)$ subgroup intact, and which maps the $T_{0k}$,
$M_{0k}$ generators into $iM_{0k}$, $iT_{0k}$ respectively ($k=1,2,3$).
In other words it exchanges the $\overline{SO}(4)$ and
$\overline{SO}(3,1)$ subgroups of the $\overline{SL}(4,R)$ group
mutually. The deunitarizing automorphism allows us to start with the
unitary (irreducible) representations of the $\overline{SL}(4,R)$ group,
and upon its application, to identify the finite (unitary)
representations of the "abstract" $\overline{SO}(4)$ compact subgroup
with nonunitary representations of the physical Lorentz group --
$\overline{SO}(3,1) = \overline{SO}(4)^{\cal A}$ . In this way, we avoid
a disease common to most of infinite-component wave equations, in
particular those based on groups containing the $\overline{SL}(4,R)$
group.

\section{Hadron Spectroscopy}

The catalogue of the $\overline{SL}(4,R)$ multiplicity-free unitary 
irreducible representations is presented in \cite{sl4} , and by making 
use of the deunitarizing automorphism ${\cal A}$, we arrive at the 
infinite-dimensional $\overline{SL}(4,R)$ representations for which 
the Lorentz subgroup is represented nonunitarily. Moreover, for the 
relevant cases the Lorentz-covariant (flat-space) infinite-component 
wave equations which determine the physical (propagating) degrees of 
freedom are given in \cite{we}, and thus we can proceed with the
actual applications to hadron classification.

In the case of mesons there are two $\overline{SL}(4,R)$ representations
at our disposal: $D_{SL(4,R)}^{ladd}(0)^{\cal A}$ and
$D_{SL(4,R)}^{ladd}(\frac{1}{2})^{\cal A}$. Having in mind the quark
model, it is most natural to classify the $\bar{q}q$ meson states
according to the $D_{SL(4,R)}^{ladd}(\frac{1}{2})^{\cal A}$
representation, i.e., to have as the lowest level the $J=0, 1$ ($^1S_{0}$
and $^3S_{1}$) states.  The $D_{SL(4,R)}^{ladd}(0)^{\cal A}$
representation would be an appropriate choice for the possible glueballs.
In the case of baryons, for the flavor  $SU(3)$  octet states we have a
unique choice of the system based on the
$[D_{SL(4,R)}^{disc}(\frac{1}{2},0) \oplus
D_{SL(4,R)}^{disc}(0,\frac{1}{2})]^{{\cal A}}$ system, while for the
decuplet states we have to make use of the symmetrized product of this
reducible representation and the finite-dimensional $\overline{SL}(4,R)$
representation $(\frac{1}{2},\frac{1}{2})$ (generalizing the Rarita -
Schwinger approach). The $\overline{SL}(4,R)$ generators have definite
space-time properties, and in particular a constrained behavior under the
parity operation: The $J_{i} = \epsilon_{ijk} M_{jk}$, $T_{ij}$, and
$T_{00}$ operators are parity even, while the $K_{i} = M_{0i}$ and $N_{i}
= T_{0i} $ are parity odd. All states of the same $\overline{SL}(3,R)$
subgroup unitary irreducible representation (Regge trajectory) thus
have the same parity; the states of an $SL(2,C) \simeq
\overline{SO}(3,1)$ or an $\overline{SO}(4)$
subgroup representation have alternating parities. For a given $SL(2,C) =
\overline{SO}(4)^{\cal A}$ representation ($j_{1},j_{2}$), the total
(spin) angular momentum is
\begin{equation}
J = J^{(1)} + J^{(2)},
\end{equation}
while the boost operator is given by
\begin{equation}
K = J^{(1)} - J^{(2)}.
\end{equation}
We find the following $J^{P}$ content of a $(j_{1},j_{2})$
$\overline{SO}(4)^{\cal A}$ representation:
\begin{equation}
J^{P} = (j_{1}+j_{2})^{P},\ (j_{1}+j_{2}-1)^{-P},\ (j_{1}+j_{2}-2)^{P},\
\cdots ,\ (|j_{1}-j_{2}|)^{\pm P}.
\end{equation}
Thus, by assigning a given parity to any state of an $\overline{SL}(4,R)$
representation, say the lowest state, the parities of all other states
are determined.

The $\overline{SL}(3,R)$ subgroup unitary irreducible representations
\cite{sl3} determine the Regge trajectory states of a given
$\overline{SL}(4,R)^{\cal A}$ representation. In decomposing an
$\overline{SL}(4,R)^{\cal A}$ representation with respect to the
$\overline{SL}(3,R)$ unitary irreducible representations, it is
convenient to use an integer quantum number $n$ that is in one-to-one 
correspondence with the $T_{00}$ operator eigen values.

The $\overline{SL}(4,R)$ ladder unitary irreducible representations 
contain an infinite sum of $\overline{SL}(3,R)$ ladder unitary
irreducible representations, i.e.,
\begin{equation*}
D_{SL(4,R)}^{ladd}(0;e_{2}) \rightarrow {\sum_{n\ even}}^{\oplus}
D_{SL(3,R)}^{ladd}(0;\sigma_{2}) \oplus {\sum_{n\ odd}}^{\oplus}
D_{SL(3,R)}^{ladd}(1;\sigma_{2}),
\end{equation*}
\begin{equation*}
D_{SL(4,R)}^{ladd}(\frac{1}{2};e_{2}) \rightarrow {\sum_{n\
even}}^{\oplus} D_{SL(3,R)}^{ladd}(1;\sigma_{2}) \oplus {\sum_{n\
odd}}^{\oplus} D_{SL(3,R)}^{ladd}(0;\sigma_{2}).
\end{equation*}
An analysis shows that the reduction of the
$D_{SL(4,R)}^{disc}(\frac{1}{2},0)$ and
$D_{SL(4,R)}^{disc}(0,\frac{1}{2})$ representations with respect to the
unitary irreducible representations of $\overline{SL}(3,R)$ is given
by the symbolic expression
\begin{equation*}
D_{SL(4,R)}^{disc}({1\over 2},0) \oplus D_{SL(4,R)}^{disc}(0,{1\over 2})
\rightarrow {\sum_{n \ even}}^{\oplus} D_{SL(3,R)}^{ladd}({1\over 2})
\oplus {\sum_{n\ odd}}^{\oplus} D_{SL(3,R)}^{disc}({3\over
2};\sigma_{2});
\end{equation*}
each $\overline{SL}(3,R)$ unitary irreducible representation appears 
infinitely many times.

We find it necessary, from a comparison with the experimental situation,
to use parity doubling, the actual spectrum displaying approximate
exchange-degeneracy features. The parity of states within an
$\overline{SL}(4,R)^{\cal A}$ representation is determined by the parity
of the lowest - $J$ state.

Thus, {\it we assign all hadron states of a given flavor to the
wave-equation-projected states corresponding to parity-doubled
$\overline{SL}(4,R)^{\cal A}$ irreducible representations} (their lowest
- $J$ states have opposite parities) \cite{hadrons}.

\vskip6pt
%%%%%%%%%%%%%%%% begin N_Sigma baryons %%%%%%%%%%%%%%%%%%
\hskip -25pt {\scriptsize
\begin{tabular}{|c|c|c|c|c|c|c|c|c|c|} \hline
\multicolumn{10}{|c|}{TABLE I \hskip20pt Assignment of $N$, $\Lambda$ and
$\Sigma$\ \ $SU(3)$ octet states} \\ \hline\hline
   \multicolumn{5}{|c|}{$D(0,\frac{1}{2})$} &
   \multicolumn{5}{|c|}{$D(0,\frac{1}{2})$} \\ \hline
$(j_{1},j_{2})$ & $J^{P}$ & $\{ N\}$ & $\{\Lambda\}$ & $\{\Sigma\}$ &
$(j_{1},j_{2})$ & $J^{P}$ & $\{ N\}$ & $\{\Lambda\}$ & $\{\Sigma\}$
\\ \hline
$(\frac{1}{2},0)$ & $\frac{1}{2}^{+}$ & $N(940)$ & $\Lambda (1116)$ &
$\Sigma (1193)$ & $(0,\frac{1}{2})$ & $\frac{1}{2}^{-}$ & $N(1535)$ &
$\Lambda (1670)$ & $\underline{\Sigma}(\sim 1500)$
\\ \hline
  & $\frac{1}{2}^{+}$ & $N(1440)$ & $\Lambda (1600)$ & $\Sigma (1660)$ &
  & $\frac{1}{2}^{-}$ & $N(1650)$ & $\Lambda (1800)$ & $\Sigma(1620)$
\\
$(\frac{3}{2},1)$ & $\frac{3}{2}^{-}$ & $N(1520)$ & $\Lambda (1690)$ &
$\Sigma (1670)$ & $(1,\frac{3}{2})$ & $\frac{3}{2}^{+}$ &
$\underline{N}(1540)$ &  $\Lambda (1890)$ & $\Sigma (1670)$
\\
  & $\frac{5}{2}^{+}$ & $N(1680)$ & $\Lambda (1820)$ &
$\Sigma (1915)$ &  & $\frac{5}{2}^{-}$ & $N(1675)$ & $\Lambda (1830)$ &
$\Sigma (1775)$
\\ \hline
  & $\frac{1}{2}^{+}$ & $N(1710)$ & $\Lambda (1800)$ & $\Sigma (1880)$ &
  & $\frac{1}{2}^{-}$ & $N(2090)$ &   & $\Sigma (1750)$
\\
  & $\frac{3}{2}^{-}$ & $N(1700)$ & $\Lambda (2000)$ &  &  &
$\frac{3}{2}^{+}$ & $N(1720)$ &  & $\Sigma (1840)$
\\
$(\frac{5}{2},2)$ & $\frac{5}{2}^{+}$ & $N(2000)$ & $\Lambda (2110)$ &  &
$(2,\frac{5}{2})$ & $\frac{5}{2}^{-}$ & $N(2200)$ &
  &
\\
  & $\frac{7}{2}^{-}$ & $N(2190)$ &  &  &  & $\frac{7}{2}^{+}$ &
$N(1990)$ & $\underline{\Lambda}(2020)$ &
\\
  & $\frac{9}{2}^{+}$ & $N(2220)$ & $\Lambda (2350)$ & $\Sigma (2455)$
  &  & $\frac{9}{2}^{-}$ & $N(2250)$ &  &
\\ \hline
  & $\frac{1}{2}^{+}$ & $\underline{N}(2100)$ &  & $\Sigma (2250)$ &
  & $\frac{1}{2}^{-}$ &  &  &
\\
  & $\frac{3}{2}^{-}$ & $N(2080)$ & $\Lambda (2325)$ &  &
  & $\frac{3}{2}^{+}$ &  &  &
\\
  & $\frac{5}{2}^{+}$ &  &  &  &  & $\frac{5}{2}^{-}$ &  &  &
\\
$(\frac{7}{2},3)$ & $\frac{7}{2}^{-}$ &  &  &  & $(3,\frac{7}{2})$
  & $\frac{7}{2}^{+}$ &  &  &
\\
  & $\frac{9}{2}^{+}$ &  &  &  &  & $\frac{9}{2}^{-}$ &  &  &
\\
  & $\frac{11}{2}^{-}$ & $N(2600)$ &  &  &  & $\frac{11}{2}^{+}$ &
  &  &
\\
  & $\frac{13}{2}^{+}$ & $N(2700)$ &  &  &  & $\frac{13}{2}^{-}$ &  &  &
\\ \hline
\end{tabular}
}
%%%%%%%%%%%%%%%%%% end N_Sigma baryons %%%%%%%%%%%%%%%%
\vskip6pt

{\bf Mesons} $(q\bar q)$: $D_{SL(4,R)}^{ladd}(\frac{1}{2};e_{2})^{\cal
A},\ \Phi ,$
\begin{equation}
\{ (j_{1},j_{2})\} = \{ ({1\over 2},{1\over 2}),\ ({3\over 2},{3\over
2}), \ ({5\over 2},{5\over 2}),\ \cdots \}
\end{equation}

{\bf Baryons} $(qqq)_{mixed symmetry}$:
$[D_{SL(4,R)}^{disc}(\frac{1}{2},0) \oplus
D_{SL(4,R)}^{disc}(0,\frac{1}{2})]^{\cal A},\ \Psi ,$
\begin{equation}
\{ (j_{1},j_{2})\} = \{ (\frac{1}{2},0),\ (\frac{3}{2},1),\
(\frac{5}{2},0),\ \cdots \} \oplus \{ (0,\frac{1}{2}),\ (1,\frac{3}{2}),\
(2,\frac{5}{2}),\ \cdots \} .
\end{equation}

{\bf Baryons} $(qqq)_{symmetric}$: $ \{
[D_{SL(4,R)}^{disc}(\frac{1}{2},0) \oplus
D_{SL(4,R)}^{disc}(0,\frac{1}{2})]^{\cal A} \otimes
D^{(\frac{1}{2},\frac{1}{2})} \}_{sym},\ \Psi_{\rho},$
\begin{equation}
\{ (j_{1},j_{2})\} = \{ (1,\frac{1}{2}), (2,\frac{3}{2}),
(3,\frac{5}{2}), \cdots \} \oplus \{ (\frac{1}{2},1), (\frac{3}{2},2),
(\frac{5}{2},3), \cdots \}.
\end{equation}

The $\overline{SO}(4)^{\cal A}$ states, when reorganized
with respect to the $\overline{SL}(3,R)$ subgroup, form an infinite sum
of Regge-type $\Delta J=2$ recurrences with the $J$ content
\begin{equation}
\{ J\} = \{ \frac{1}{2},\ \frac{5}{2},\ \frac{9}{2},\ \cdots \}, \{ J\} =
\{ \frac{3}{2},\ \frac{7}{2},\ \frac{11}{2},\ \cdots \}.
\end{equation}
The former states belong to $D_{SL(3,R)}^{ladd}(\frac{1}{2})$, while the
latter ones are projected out of $D_{SL(3,R)}^{ladd}(\frac{3}{2},
\sigma_{2})$ by the field equations.  Note that we have thus achieved the
goal of a fully relativistic algebraic model in terms of the total
angular momentum $J$.

\vskip6pt
%%%%%%%%%%%%%%%% begin Delta_Sigma baryons %%%%%%%%%%%%%%
{\scriptsize
\begin{tabular}{|c|c|c|c|c|c|c|c|} \hline
\multicolumn{8}{|c|}{TABLE II \hskip20pt Assignment of $\Delta$ and
$\Sigma$\ \ $SU(3)$ decuplet states. } \\ \hline\hline
  \multicolumn{4}{|c|}{$D(\frac{1}{2},0)_{\mu}$} &
  \multicolumn{4}{|c|}{$D(0,\frac{1}{2})_{\mu}$}
\\ \hline
$(j_{1},j_{2})$ & $J^{P}$ & $\{ \Delta \}$ & $\{\Sigma\}$ &
$(j_{1},j_{2})$ & $J^{P}$ & $\{ \Delta \}$ & $\{\Sigma\}$
\\ \hline
$(1,\frac{1}{2})$ & $\frac{1}{2}^{-}$ & $\Delta (1620)$ &
  & $(\frac{1}{2},1)$ & $\frac{1}{2}^{+}$ & $\underline{\Delta}(1550)$  &
$\Sigma (1770)$
\\
  & $\frac{3}{2}^{+}$ & $\Delta (1232)$ & $\Sigma (1385)$
  & & $\frac{3}{2}^{-}$ & $\Delta (1700)$  & $\Sigma (1580)$
\\ \hline
  & $\frac{1}{2}^{-}$ & $\Delta (1900)$  & $\underline{\Sigma} (2000)$ &
  & $\frac{1}{2}^{+}$ & $\Delta (1910)$  &
\\
$(2,\frac{3}{2})$ & $\frac{3}{2}^{+}$ & $\Delta (1600)$  & $\Sigma
(1690)$ & $(\frac{3}{2},2)$ & $\frac{3}{2}^{-}$ & $\underline{\Delta
}(1940)$ & $\Sigma (1940)$
\\
  & $\frac{5}{2}^{-}$ &  &  &  & $\frac{5}{2}^{+}$ &
$\Delta (1905)$ &
\\
  & $\frac{7}{2}^{+}$ & $\Delta (1950)$  & $\Sigma (2030)$ &  &
$\frac{7}{2}^{-}$ &  &
\\ \hline
  & $\frac{1}{2}^{-}$ & $\underline{\Delta}(2150)$ &  &  &
$\frac{1}{2}^{+}$ &  &
\\
  & $\frac{3}{2}^{+}$ & $\Delta (1920)$  & $\Sigma (2080)$ &  &
$\frac{3}{2}^{-}$ &  &
\\
$(3,\frac{5}{2})$ & $\frac{5}{2}^{-}$ & $\Delta (1930)$ &  &
$(\frac{5}{2},3)$ & $\frac{5}{2}^{+}$ & $\Delta (2000)$ &
$\underline{\Sigma}(2070)$
\\
  & $\frac{7}{2}^{+}$ &  &  &  & $\frac{7}{2}^{-}$ &
$\underline{\Delta}(2200)$  & $\underline{\Sigma}(2150)$
\\
  & $\frac{9}{2}^{-}$ & $\Delta (2400)$  & & & $\frac{9}{2}^{+}$ &
$\Delta (2300)$   &
\\
  & $\frac{11}{2}^{+}$ & $\Delta (2420)$  & $\Sigma (2620)$ &  &
$\frac{11}{2}^{-}$ &  &
\\ \hline
  & $\frac{1}{2}^{-}$ &  &  &  & $\frac{1}{2}^{+}$ &  &
\\
  & $\frac{3}{2}^{+}$ &  &  &  & $\frac{3}{2}^{-}$ &  &
\\
  & $\frac{5}{2}^{-}$ & $\underline{\Delta}(2350)$ &  &  &
$\frac{5}{2}^{-}$ &  &
\\
$(4,\frac{7}{2})$ & $\frac{7}{2}^{+}$ & $\Delta (2390)$ &  &
$(\frac{7}{2},4)$ & $\frac{7}{2}^{-}$ &  &
\\
  & $\frac{9}{2}^{-}$ &  &  &  & $\frac{9}{2}^{+}$ &  &
\\
  & $\frac{11}{2}^{+}$ &  &  &  & $\frac{11}{2}^{-}$ &  &
\\
  & $\frac{13}{2}^{-}$ & $\Delta (2750)$ &  &  & $\frac{13}{2}^{+}$ & &
\\
  & $\frac{15}{2}^{+}$ & $\Delta (2950)$ &  &  & $\frac{15}{2}^{-}$ &  &
\\ \hline
\end{tabular}
}
%%%%%%%%%%%%%%%%%% end Delta_Sigma baryons %%%%%%%%%%%%%%%%
\vskip6pt

As an example we present in the Table I the $N$, $\Lambda$, and $\Sigma$
octet states $\{ {\bf 8}\}$, while the $\Delta$ and $\Sigma$ decuplet
states $\{ {\bf 10}\}$ are presented in Table II. The $SU(3)$ $\Sigma$
assignment is not known completely. Note that the $J=\frac{1}{2}$ $\{
{\bf 10}\}$ states come from the $J=0$ part of the
$(\frac{1}{2},\frac{1}{2})$ explicit index in $\Psi_{\mu}$ of (C9), while
the other $\{ {\bf 10}\}$ states come from the $J=1$ part - thus a
discrepancy in mass.

We find a striking match between the ($J^{P}$, mass) values and the wave
- equation - projected $\overline{SL}(4,R)^{\cal A}$ representation
states. Moreover, a remarkably simple mass formula (straightforward
generalization of the mass-spin - Regge relation) fits these infinite
systems of hadronic states.  For the $\{ {\bf 8}\}$ and the higher-spin
$\{ {\bf 10}\}$ baryon resonances we write:
\begin{equation}
m^{2} = m_{0}^{2} + (\alpha_{f}^{\prime})^{-1} (j_{1} +j_{2} -
\frac{1}{2} - \frac{1}{2}n),
\end{equation}
where $m_{0}$ is the mass of the lowest-lying state,
$\alpha_{f}^{\prime}$ is the slope of the Regge trajectory for that
flavor. The linear $J\ \simeq m^2$ relation is taken here
phenomenologically. However, it will be demonstrated below that this
relation can indeed be derived from Chromogravity dynamics with a
natural choice of a Lagrangian.

\section{$J \simeq m^2$ relation}

In the absence of Chromogravity, the matter Lagrangian would be
\cite{Jm2}
\begin{equation}
L_{M} = \overline{\Psi} i X^{\mu}{\partial}_{\mu} \Psi + {\partial}^{\mu}
\Phi {\partial}_{\mu} \Phi ,
\end{equation}
invariant under global $\overline{SL}(4,R).$ The Hilbert spaces of $\Psi$
and $\Phi$ are given by the representations of $\overline{SA}(4,R)$.
Chromogravity enters through the replacement ${\partial}_{\mu}
\rightarrow {\hat D}_{A}$, where the index "$A$" denotes a local frame:
${\hat D}_{\mu} = {\partial}_{\mu} - {{\Gamma}^{A}}_{B\mu} {Q_{A}}^{B}$,
${\hat D}_{A} = {e_{A}}^{\mu}{\hat D}_{\mu}$ with $\Gamma$ the
connection and ${e_{A}}^{\mu} \cdot {e_{\mu}}^{B} =
{{\delta}_{A}}^{B}$, $e$ the chromogravity tetrad; ${Q_{A}}^{B}$ is the
$sl(4,R)$ algebraic generator in the tangent frame. We use ${\hat D}$ for
the full covariant derivative with $sl(4,R)$ connection.
${e^{A}}_{\mu}(x)$ and ${{\Gamma}^{A}}_{B\mu}(x)$ can be taken as gauge
fields for $\overline{SA}(4,R)$:
\begin{equation}
{\delta}_{(\epsilon ,\alpha )}\Psi = [ - {\epsilon}^{A}(x){\partial}_{A}
- {{\alpha}^{A}}_{B}(x) {V_{A}}^{B}]\Psi.
\end{equation}

As in gravity, the corresponding field strengths are the torsion and the
(generalized) curvature \cite{mag}, i.e.
\begin{eqnarray}
&&{{\hat R}^{A}}\,_{\mu\nu} = {\partial}_{\mu}{e^{A}}_{\nu} +
{{\Gamma}^{A}}_{B\mu}{e^{B}}_{\nu} - (\mu \leftrightarrow \nu ) \\
&&{{\hat R}^{A}}\,_{B\mu\nu} = {\partial}_{\mu}{{\Gamma}^{A}}_{B\nu} +
{{\Gamma}^{C}}_{B\mu}{{\Gamma}^{A}}_{C\nu} - (\mu \leftrightarrow \nu )
\end{eqnarray}
\noindent The Noether currents resulting from this
$\overline{SA}(4,R)$ invariance are the energy-momentum and
hypermomentum,
\begin{eqnarray}
&&{{\Theta}_{A}}^{\mu} = \frac{1}{\hat e}\  \frac{\delta L_{M}}{\delta
{e^{A}}_{\mu}}, \hskip20pt {\hat e} \equiv det({e^{A}}_{\mu}),\\
&&{{\Upsilon}^{B}}_{A\mu} = \frac{1}{\hat e}\  \frac{\delta L_{M}}{\delta
{{\Gamma}^{A}}_{B\mu}},
\end{eqnarray}
\noindent with the symmetric
$(AB)$ pairs denoting shear currents and the antisymmetric pairs $[AB]$
representing angular momentum.

The effective action for this IR (zero-color) hadron sector of QCD,
written as a Chromogravitational theory, with matter in
$\overline{SL}(4,R)$ manifields, then becomes
\begin{equation}
I = \int d^{4}x\sqrt{-G}\bigl\{ -a R_{\mu\nu}R^{\mu\nu} + b R^{2} -
cl_{G}^{-2} R + l_{S}^{-2} {\Sigma}_{\mu\nu}^{\ \
\sigma}{\Sigma}^{\mu\nu}_ {\ \ \sigma} + l_{Q}^{-2} {\Delta}_{\mu\nu}^{\
\ \sigma} {\Delta}^{\mu\nu}_{\ \ \sigma} + L_{M} \bigr\} .
\end{equation}

The first three terms constitute the Lagrangian that yields the $p^{-4}$
propagators.  The fourth and fifth terms are spin-spin and shear-shear
contact interaction terms.

We linearize the theory in terms of $H_{\mu\nu}(x) = G_{\mu\nu}(x) -
{\eta}_{\mu\nu}$, where ${\eta}_{\mu\nu}$ is the Minkowski metric. Taking
just the homogeneous part, as required for the evaluation of the
propagator, we get for the $H_{\mu\nu}$ field the equation of motion
\begin{equation}
\bigl( \frac{a}{4}\ {\square}^{\ 2} - \frac{1}{2}l_{S}^{-2}\
{\Sigma}_{\rho\eta}^{\ \ \lambda}{\Sigma}^{\rho\eta}_{\ \ \lambda} -
\frac{1}{2}l_{Q}^{-2}\ {\Delta}_{\rho\eta}^{\ \ \lambda}
{\Delta}^{\rho\eta}_{\ \ \lambda}\bigr) H_{\mu\nu} = 0,
\end{equation}
\noindent which becomes in momentum space
\begin{equation}
\bigl( \frac{a}{4}\ {(p^{2})}^{2} - \frac{1}{2} l_{S}^{-2} f_{S}\
M_{\eta}^{\ \lambda}M^{\eta}_{\ \lambda} - \frac{1}{2} l_{Q}^{-2} f_{Q}\
T_{\eta}^{\ \lambda}T^{\eta}_{\ \lambda} \bigr) H_{\mu\nu}(p) =0.
\end{equation}

For pseudo-gravity, we may regard these equations as {\it the dynamical
equations above the theory's "vacuum", as represented by hadron matter
itself}.

In the rest frame (stability) "little" group is $\overline{SL}(3,R)
\subset \overline{SL}(4,R)$. Taking a hadron's rest frame
\begin{equation}
M_{\eta}^{\ \lambda}M^{\eta}_{\ \lambda} \rightarrow M_{i}^{\ j}M^{i}_{\
j} \rightarrow J(J + 1),
\end{equation}
\begin{equation}
T_{\eta}^{\ \lambda}T^{\eta}_{\ \lambda} \rightarrow T_{i}^{\ j}T^{i}_{\
j} \rightarrow M_{i}^{\ j}M^{i}_{\ j} - A^{2}_{sl(3,R)} \rightarrow J(J +
1) - C^{2}_{sl(3,R)},
\end{equation}
\noindent where $C^{2}$ is the $sl(3,R)$ quadratic invariant.

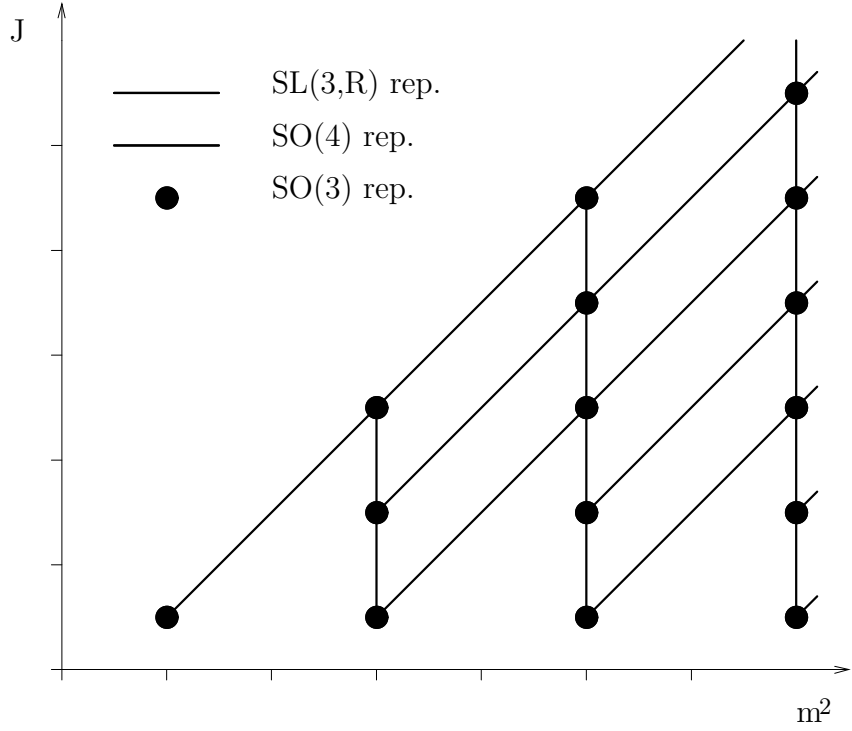
\begin{center}
\setlength{\unitlength}{0.00061242in}
\begingroup\makeatletter\ifx\SetFigFont\undefined%
\gdef\SetFigFont#1#2#3#4#5{%
  \reset@font\fontsize{#1}{#2pt}%
  \fontfamily{#3}\fontseries{#4}\fontshape{#5}%
  \selectfont}%
\fi\endgroup%
{\renewcommand{\dashlinestretch}{30}
\begin{picture}(7212,6192)(0,-10)
\thicklines
\put(1350,900){\blacken\ellipse{180}{180}}
\put(1350,900){\ellipse{180}{180}}
\put(3150,900){\blacken\ellipse{180}{180}}
\put(3150,900){\ellipse{180}{180}}
\put(3150,1800){\blacken\ellipse{180}{180}}
\put(3150,1800){\ellipse{180}{180}}
\put(3150,2700){\blacken\ellipse{180}{180}}
\put(3150,2700){\ellipse{180}{180}}
\put(4950,900){\blacken\ellipse{180}{180}}
\put(4950,900){\ellipse{180}{180}}
\put(4950,1800){\blacken\ellipse{180}{180}}
\put(4950,1800){\ellipse{180}{180}}
\put(4950,2700){\blacken\ellipse{180}{180}}
\put(4950,2700){\ellipse{180}{180}}
\put(4950,3600){\blacken\ellipse{180}{180}}
\put(4950,3600){\ellipse{180}{180}}
\put(4950,4500){\blacken\ellipse{180}{180}}
\put(4950,4500){\ellipse{180}{180}}
\put(6750,900){\blacken\ellipse{180}{180}}
\put(6750,900){\ellipse{180}{180}}
\put(6750,1800){\blacken\ellipse{180}{180}}
\put(6750,1800){\ellipse{180}{180}}
\put(6750,2700){\blacken\ellipse{180}{180}}
\put(6750,2700){\ellipse{180}{180}}
\put(6750,3600){\blacken\ellipse{180}{180}}
\put(6750,3600){\ellipse{180}{180}}
\put(6750,4500){\blacken\ellipse{180}{180}}
\put(6750,4500){\ellipse{180}{180}}
\put(6750,5400){\blacken\ellipse{180}{180}}
\put(6750,5400){\ellipse{180}{180}}
\put(1350,4500){\blacken\ellipse{180}{180}}
\put(1350,4500){\ellipse{180}{180}}
\thinlines
\path(450,5850)(450,360)
\path(360,450)(6750,450)
\path(450,1350)(360,1350)
\path(450,2250)(360,2250)
\path(450,3150)(360,3150)
\path(450,4050)(360,4050)
\path(450,4950)(360,4950)
\path(1350,450)(1350,360)
\path(2250,450)(2250,360)
\path(3150,450)(3150,360)
\path(4050,450)(4050,360)
\path(4950,450)(4950,360)
\path(5850,450)(5850,360)
\thicklines
\path(1350,900)(6300,5850)
\path(3150,900)(6930,4680)
\path(3150,1800)(6930,5580)
\path(4950,900)(6930,2880)
\path(4950,1800)(6930,3780)
\path(6750,900)(6930,1080)
\path(6750,1800)(6930,1980)
\thinlines
\path(6750,450)(7200,450)
\path(7080.000,420.000)(7200.000,450.000)(7080.000,480.000)
\path(450,5850)(450,6165)
\path(480.000,6045.000)(450.000,6165.000)(420.000,6045.000)
\thicklines
\path(3150,900)(3150,2700)
\path(4950,900)(4950,4500)
\path(6750,900)(6750,5850)
\path(900,5400)(1800,5400)
\path(900,5400)(1800,5400)
\path(900,4950)(1800,4950)
\path(900,4950)(1800,4950)
\put(2250,5400){\makebox(0,0)[lb]{\smash{{{\SetFigFont{12}{14.4}{\rmdefault}{\mddefault}{\updefault}SL(3,R) rep.}}}}}
\put(2250,4950){\makebox(0,0)[lb]{\smash{{{\SetFigFont{12}{14.4}{\rmdefault}{\mddefault}{\updefault}SO(4) rep.}}}}}
\put(2250,4500){\makebox(0,0)[lb]{\smash{{{\SetFigFont{12}{14.4}{\rmdefault}{\mddefault}{\updefault}SO(3) rep.}}}}}
\put(0,5850){\makebox(0,0)[lb]{\smash{{{\SetFigFont{12}{14.4}{\rmdefault}{\mddefault}{\updefault}J}}}}}
\put(6750,0){\makebox(0,0)[lb]{\smash{{{\SetFigFont{12}{14.4}{\rmdefault}{\mddefault}{\updefault}m}}}}}
\put(6945,71){\makebox(0,0)[lb]{\smash{{{\SetFigFont{9}{10.8}{\rmdefault}{\mddefault}{\updefault}2}}}}}
\end{picture}
}

\end{center}

As a result, we find that in a rest frame, all hadronic states belonging
to a single $\overline{SL}(3,R)$ (unitary) irreducible representation
(i.e. one value of $C^{2}_{sl(3,R)}$) lay on a single trajectory in the
Chew-Frautschi plane, i.e.
\begin{equation}
(J + \frac{1}{2})^{2} = ({\alpha '}m^{2})^{2} + {\alpha}_{0}^{2},
\end{equation}
\begin{equation}
(\alpha ')^{2} = \bigl[ \frac{2}{a} (l_{S}^{-2} f_{S} + l_{Q}^{-2}
f_{Q})\bigr]^{-1},
\end{equation}
\begin{equation}
{\alpha}_{0}^{2} = \frac{1}{4} + \frac{l_{Q}^{-2}f_{Q}}{l_{S}^{-2} f_{S} +
l_{Q}^{-2}f_{Q}} C^{2}_{sl(3,R)}.
\end{equation}
\noindent $\alpha '$ is the (asymptotic) trajectory slope, $S$ is the
Cartan's chromo-torsion tensor, while $Q = D\ G$ is the
chromo-non-metricity tensor. Neglecting a slight bending at small
$m^{2}$,i.e. the ${\alpha}_{0}^{2}$ term, we finally obtain the linear
Regge trajectory
\begin{equation}
J = {\alpha '} m^{2} - \frac{1}{2}.
\end{equation}
A combined result of the $\overline{SL}(4,R) \supset R_{+}\otimes
\overline{SL}(3,R)$ representation reduction states and the $J \simeq
m^2$ relation is illustrated on the above figure.

\section{IBM -- Interacting Boson Model Derivation}

The Interacting Boson Model has been very successful as a dynamical
symmetry in correlating as well as providing an understanding of a large
amount of data which manifest the collective behavior of nuclei. The
model's point of departure is the observation that the two lowest levels
in the great majority of even-even nuclei are the $0^{+}$ and $2^{+}$
levels, with relatively close excitation energies, realized by proton or
neutron pairs. The model postulates a corresponding phenomenological
$U(6)$ symmetry.

As demonstrated above, the strongly-coupled IR region in QCD is
approximated by the exchange of a phenomenological chromometric
di-gluon field $G_{\mu\nu}(x)$. The $G_{\mu\nu}(x)$ acts formally
as a Riemannian metric, i.e. it obeys the following Riemannian constraint:
\begin{equation}
D_{\sigma}G_{\mu\nu}(x) = 0.
\end{equation}
where $D_{\sigma}$ is the covariant derivative of the effective gravity,
with the connection given by a Christoffel symbol constructed with this
effective metric. As a result, the surviving quanta are color neutral
and have  $J^P = 0^+,\  2^+$, with symmetric couplings to matter fields.

We now stretch the Chromogravity application from a single hadron case
over to the composite hadronic system of nuclear matter \cite{ibm} in a
Van der Waals fashion. As in the hadronic case, the next vibrational,
rotational or pulsed excitations will correspond to the "addition" of one
such collective  color-singlet multigluon quantum superposition, while
the basic exchanged quantum is generated by "gluonium" $G_{\mu\nu}(x)$.

An effective Riemannian metric induces the corresponding Einsteinian
dynamics. The invariant action in which the Einstein-like Lagrangian $R$
is accompanied by a parametrized combination of the allowed quadratic
terms reads
\begin{equation}
I_{inv} = - \int d^{4}x \sqrt{-G} (\alpha R_{\mu\nu} R^{\mu\nu} - \beta
R^{2} + \gamma{\kappa}^{-2} R).
\end{equation}

The theory is renormalizable, a feature befitting the present
application, since QCD is renormalizable, but is not unitary, which also
befits this application: a "piece" of QCD should not be unitary,
considering that QCD is an irreducible theory. The renormalizability is
caused by $p^{-4}$ propagators. {\it $p^{-4}$ propagators are dynamically
equivalent to confinement}!

Moreover, it has been shown that the presence of the quadratic terms in
the action  induces a potential $\sim \frac{1}{r} + r + r^{2}$.

{\bf Nuclei}. Out of the $10$ components of $G_{\mu\nu}$ the 6 that survive
the 4 Riemannian constraints have spin/parity assignments $J^{P} = 0^{+},
\ 2^{+}$.

The non-relativistic subgroup of $SL(4,R)$ is $SL(3,R)$. Under this
group, the $0^{+}$ and $2^{+}$ states span together one irreducible
6-dimensional representation, thus both $0^{+}$ and $2^{+}$ couple with
the same strength to nucleons. There is thus full justification for the
IBM postulate of a $U(6)$ symmetry between the defining states!

The closed shells assume the role of "vacua", as rigid structures.
Gluonium excitations should then be searched for in the valence nucleon
systematics.

The corresponding Hamiltonian in terms of $b = \{s,d\}$ and $b^{+} =
\{s^{+}, d^{+}\}$ that represent the destruction and creation of a
6-dimensional gluonium quantum reads:

\begin{equation*}
H = \frac{1}{M^{3}} \int dk \{ C_{1} \frac{k^2}{\kappa^2} (b^{+}b) + C_{2}
\frac{k^2}{\kappa^2}(b^{+}\bullet b)(b^{+}\bullet b) + A_{1}
k^{4}(b^{+}\bullet b)(b^{+}\bullet b)
\end{equation*}
\begin{equation}
+ A_{2} k^{4} (b^{+}\bullet b)(b^{+}\bullet b)(b^{+}\bullet b) + A_{3}
k^{4} (b^{+}\bullet b)(b^{+}\bullet b) (b^{+}\bullet b)(b^{+}\bullet b)).
\end{equation}

{\bf Symmetries of deformed nuclei}

 In the quantum case, we can write,
$G_{\mu\nu} = T_{\mu\nu} + U_{\mu\nu}$, where
\begin{equation}
T_{\mu\nu} = \eta_{ab} \int d{\tilde k}
[{\alpha^{a}_{\mu}}^{+}(k){\alpha^{b}_{\nu}}^{+}(k) e^{2ikx} +
\alpha^{a}_{\mu}(k)\alpha^{b}_{\nu}(k) e^{-2ikx}],
\end{equation}
and
\begin{equation}
U_{\mu\nu} = \eta_{ab} \int d{\tilde k} [{\alpha^{a}_{\mu}}^{+}(k)
\alpha^{b}_{\nu}(k) + \alpha^{a}_{\mu}(k) {\alpha^{b}_{\nu}}^{+}(k)].
\end{equation}

This time we use the creation and annihilation operators
${\alpha^{a}_{\mu}}^{+}, \alpha^{b}_{\nu}$ of the QCD gluon itself, which
can be regarded somewhat like a tetrad field with respect to $G_{\mu\nu}$
as a metric. For this to fit the formalism, we have to separate out the
"rigid" piece (analogous to $e^{i}_{\mu} = {\delta}^{i}_{\mu} +
h^{i}_{\mu}$ in the tetrad case).  Here this is the "flat connection"
$N^{a}_{\mu}$, i.e. the zero-mode of the field.

The operators $T_{\mu\nu}$ and $U_{\mu\nu}$, together with the operators
\begin{equation} S_{\mu\nu} =   \eta_{ab} \int d{\tilde k}
[{\alpha^{a}_{\mu}}^{+}(k) \alpha^{b}_{\nu}(k) - \alpha^{a}_{\mu}(k)
{\alpha^{b}_{\nu}}^{+}(k)],
\end{equation}
close respectively on the algebras of $GL(4,R)$ and $U(1,3)$.  Note that
the largest (linearly realized) algebra with generators quadratic in the
${\alpha_{\mu}}^{+}, \ \alpha_{\mu}$ operators is the algebra of
$Sp(4,R)$. This algebra contains both previous ones:
\begin{equation}
Sp(4,R) \supset \left\{ \begin{matrix} U(1,3) & \supset & SU(1,3) \\
                 GL(4,R) & \supset & SL(4,R) \\
                 T_{10}\wedge SO(1,3) & \supset & T_{9}\wedge SO(1,3)
                 \end{matrix}\right\} \supset SO(1,3)
\end{equation}
The $GL(4,R)$ algebra represents a Spectrum-Generating Algebra for the
set of hadron states of a given flavor. In the case of $U(1,3)$, when
selecting a time-like vector (for massive states), the stability subgroup
is $U(3)$, a compact group with finite representations -- as against the
non-compact $SL(3,R)$ for $SL(4,R)$. This fits with a situation in nuclei
in which the symmetries are physically realized over pairs of "valency"
nucleons outside of closed shells, as in the case of IBM: there is a
finite number of such pairs, and the excitations thus have to fit within
finite representations.

\end{document}